\documentclass[conference]{IEEEtran}
\IEEEoverridecommandlockouts

\usepackage{cite}
\usepackage{amsmath,amssymb,amsfonts}
\usepackage{algorithmic}
\usepackage{tcolorbox}
\usepackage{graphicx}
\usepackage{textcomp}
\usepackage{xcolor}
\usepackage{booktabs}
\usepackage{multirow}
\usepackage{hyperref}
\usepackage{tcolorbox}
\usepackage{tabularx}

\def\BibTeX{{\rm B\kern-.05em{\sc i\kern-.025em b}\kern-.08em
    T\kern-.1667em\lower.7ex\hbox{E}\kern-.125emX}}
\begin{document}

\title{Multi-Agent Debate Strategies to Enhance Requirements Engineering with Large Language Models\\
}

\author{
\IEEEauthorblockN{Marc Oriol}
\IEEEauthorblockA{\textit{Universitat Politècnica de Catalunya}\\
Barcelona, Spain \\
marc.oriol@upc.edu}
\and

\IEEEauthorblockN{Quim Motger}
\IEEEauthorblockA{\textit{Universitat Politècnica de Catalunya}\\
Barcelona, Spain \\
joaquim.motger@upc.edu}
\and

\IEEEauthorblockN{Jordi Marco}
\IEEEauthorblockA{\textit{Universitat Politècnica de Catalunya}\\
Barcelona, Spain \\
jordi.marco@upc.edu}
\and

\IEEEauthorblockN{Xavier Franch}
\IEEEauthorblockA{\textit{Universitat Politècnica de Catalunya}\\
Barcelona, Spain \\
xavier.franch@upc.edu}
}

\maketitle

\begin{abstract}
Context: Large Language Model (LLM) agents are becoming widely used for various Requirements Engineering (RE) tasks. Research on improving their accuracy mainly focuses on prompt engineering, model fine-tuning, and retrieval augmented generation. However, these methods often treat models as isolated black boxes - relying on single-pass outputs without iterative refinement or collaboration, limiting robustness and adaptability.
Objective: We propose that, just as human debates enhance accuracy and reduce bias in RE tasks by incorporating diverse perspectives, different LLM agents debating and collaborating may achieve similar improvements. Our goal is to investigate whether Multi-Agent Debate (MAD) strategies can enhance RE performance.
Method: We conducted a systematic study of existing MAD strategies across various domains to identify their key characteristics. To assess their applicability in RE, we implemented and tested a preliminary MAD-based framework for RE classification.
Results: Our study identified and categorized several MAD strategies, leading to a taxonomy outlining their core attributes. Our preliminary evaluation demonstrated the feasibility of applying MAD to RE classification.
Conclusions: MAD presents a promising approach for improving LLM accuracy in RE tasks. This study provides a foundational understanding of MAD strategies, offering insights for future research and refinements in RE applications.
\end{abstract}


\section{Introduction}
The emergence of generative Artificial Intelligence (AI) and Large Language Models (LLMs) has revolutionized the development of AI agents, which are increasingly being applied across various domains of Software Engineering, including Requirements Engineering (RE)~\cite{Lo2024}. These AI agents can assist with a wide range of RE activities and tasks, including requirements elicitation, analysis, classification, ambiguity detection, negotiation, and validation, among others~\cite{Arora2024}.

However, for these AI agents to be trustworthy and operate autonomously, it is crucial that they demonstrate high-quality performance across several dimensions, including but not limited to accuracy, explainability, robustness and fairness. Regarding accuracy, which is the main focus of this paper, AI agents commonly face several challenges when performing RE tasks. These include hallucinations, lack of domain-specific language, bias in training data, and ambiguity, to name a few~\cite{naveed2024}.

Research on improving accuracy has focused mainly on providing novel approaches based on prompt engineering~\cite{Ronanki2024}, model fine-tuning~\cite{Liu2024}, and retrieval augmented generation (RAG)~\cite{Arora2024b}.
Although prompt engineering has been a key technique until now, recent studies suggest that its effectiveness is diminishing in newer generative AI models, such as GPT-4o and OpenAI o3~\cite{wang2024}. Similarly, model fine-tuning and RAG are often constrained by the need for large volumes of high-quality domain-specific data, which may not always be feasible to obtain.

This raises the key question of what techniques can be employed to further enhance the accuracy of AI agents based on the latest AI models and in the absence of further domain-specific data. A novel approach addressing this challenge is inspired from human collaboration: in order to improve accuracy, reduce bias, and enhance performance, humans collaborate by giving their opinions, debating discrepancies and seeking consensus. Analogously, different AI agents working and debating collaboratively may achieve similar improvements. This emerging paradigm, known as Multi-Agent Debate (MAD)~\cite{p28-chan-2024, p04-liang-2024}, has gained popularity with the advent of advanced LLMs. MAD has already been applied in a wide range of tasks, from solving complex mathematical problems~\cite{p03-li-2024} to improving the accuracy of general knowledge tests~\cite{p17-wang-2025}. However, the results of a systematic mapping we are conducting (cf. Section~\ref{sec:MAD_stra}) show that no MAD approach has ever been proposed to support RE. 

To address this gap, we explore the use of MAD to support RE tasks. In particular, in this research preview, we (1) examine the key characteristics of MAD strategies found in the literature across multiple domains, and (2) assess their applicability to RE tasks by implementing a representative case as a proof-of-concept. This two-fold objective yields the following two research questions:

\textbf{RQ1:} What MAD strategies have been proposed in the literature, and what are the key characteristics of their debate mechanisms?

\textbf{RQ2:} To which degree can these MAD strategies be applied in the field of RE and improve the accuracy of RE tasks compared to single AI agents? 

\section{MAD strategies}
\label{sec:MAD_stra}

To answer RQ1, we conducted a first iteration of a systematic mapping study, following the guidelines of Wohlin~\cite{Wohlin2014} combined with the search strategies of Kitchenham et al.~\cite{keele2007}, to obtain a set of seed papers over which we performed snowballing. Below, we present the methodological details and the current preliminary results. 

\subsection{Search methodology}
\label{sec:search-methodology}
We defined a search string that captured the concept of Multi-Agent Debate on one side and LLMs or AI agents on the other, incorporating both singular and plural forms as well as relevant acronyms. After some piloting, the final search string was written as: 

\begin{tcolorbox}
(MAD OR ``Multi-Agent Debate") AND (LLM OR ``Large Language Model" OR LLMS OR ``Large Language Models" OR ``AI Agent" OR ``AI Agents")
\end{tcolorbox}

The search was conducted using the Scopus database, over title, abstract and keywords of papers; no restriction was set in terms of domain or date of publication. 

It is important to note that the goal of this search was not to retrieve all available papers on MAD but rather to obtain a relevant list of high-quality papers to be used as the seed for the subsequent snowballing process. As such, we prioritized precision over recall and deliberately chose not to include additional alternative terms for ``MAD" or ``Multi-Agent Debate." as these alternative terms (e.g. "debate", "discussion") introduced substantial noise with limited relevant results. Likewise, we selected Scopus for its broad coverage of high-quality, peer-reviewed research across disciplines~\cite{Martín-Martín2021}.

The inclusion and exclusion criteria for the selection of papers are presented in Table~\ref{tab:criteria}. 
The selection process was carried out first by inspecting their titles, then their abstract, and finally a rapid review of the full text.

\begin{table}[h]
    \centering
    \renewcommand{\arraystretch}{1.2}
    \setlength{\tabcolsep}{3pt} 
    \caption{Inclusion and Exclusion Criteria}
    \label{tab:criteria}
\begin{tabular}{@{}p{4cm}p{4cm}@{}}
\toprule
\textbf{Inclusion Criteria} & \textbf{Exclusion Criteria} \\
\midrule
Studies that propose or use one or more debate strategies for multiple AI agents conducting the same task. & 
The paper does not provide enough details on the strategy applied for the debate (e.g., the debate is not the focus of the paper). \\
\cmidrule(l){2-2}
& Debates are among AI agents and humans, rather than among AI agents only. \\
\cmidrule(l){2-2}
& Debates are not on natural language tasks (e.g., debates on image recognition). \\
\bottomrule
\end{tabular}
\end{table}

The resulting list of papers served as the seed set for the snowballing process, which includes both backward and forward snowballing. This snowballing process is being conducted iteratively in multiple rounds until saturation is reached. This snowballing process helps to achieve two goals: (1) capturing not only the papers satisfying our query, but also broader relevant works that meet our inclusion/exclusion criteria; and (2) capturing relevant papers available only on arXiv, which today is a fundamental source of timely research in the RE discipline that cannot be disregarded in a literature study. In this research preview, we present our initial findings after completing the first iteration of backward snowballing. Details of the search process and data analysis are available in our replication package~\cite{Oriol2025}.

\subsection{Search results}
The search string in Scopus delivered 29 results. Up to 17 of them either did not meet the inclusion criteria or met one or more exclusion criteria, leading to 12 seed papers. The backward snowballing process was applied first, resulting in 527 additional papers; after removing 169 duplicated papers, 358 papers remained to review. From them, 13 papers satisifed the IC/EC criteria, leading to a total of 25 papers. If multiple versions of the same paper were published on arXiv, we selected the most recent version. When a paper appeared both on arXiv and in a conference or journal, we prioritized the conference or journal version for our analysis.

From the total of 25 publications obtained, the earliest publications appeared in 2023, with a total of 4 papers (8 if original arXiv versions that were later published in conferences or journals are included). In 2024, there were 17 publications. As of March 10th 2025 (the date when the search was conducted), we have identified 4 additional papers.

Notably, we did not find any paper applying MAD in Requirements Engineering (RE) nor in the Software Engineering domain.

\subsection{Analysis}
\label{sec:search-analysis}

\begin{figure*}[htbp]
  \centering
  \includegraphics[width=\textwidth]{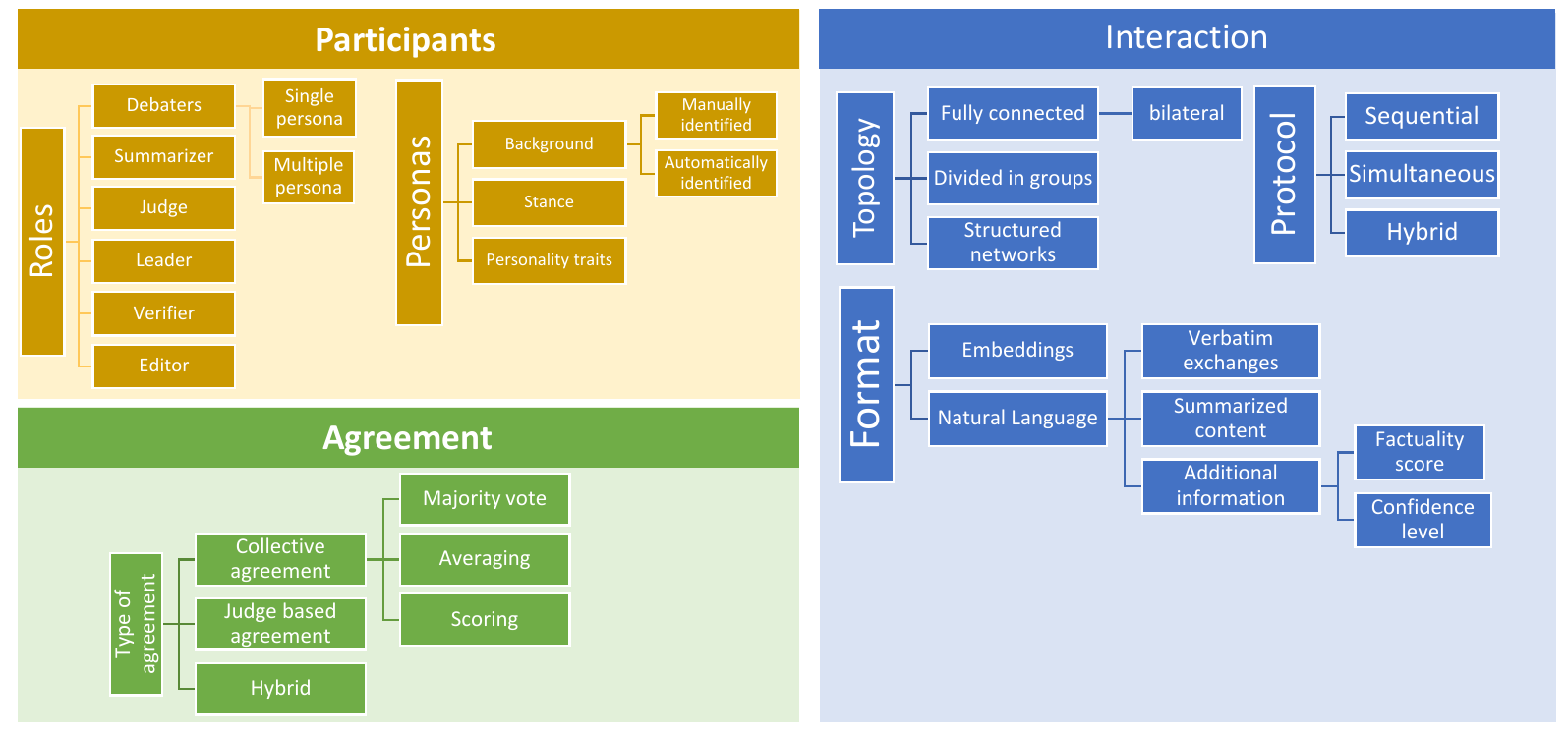}
  \caption{Taxonomy of Multi-Agent Debate characteristics}
  \label{fig:taxonomy}
\end{figure*}

Applying feature extraction to the selected papers, we identified a number of codes that characterize the different methodologies used in MAD. These codes fall into three primary categories: the \textit{participants} involved, the \textit{interactions} held during the debate, and the strategies for reaching \textit{agreement}. A taxonomy representing the coding schema is presented in Figure~\ref{fig:taxonomy}. Below, we describe how the papers in our review implement MAD in accordance to the defined taxonomy. It is important to note that a single paper may include multiple variations or approaches, and therefore, it may be represented under more than one aspect of the taxonomy.

\subsubsection{Participants}
We identify two sub-categories related to participants: roles involved and characterization of personas.

\textbf{Roles involved}:  \textit{Debaters}, defined as AI agents that present positions based on their prompt configurations and engage in the debate to reach a consensus, are the basic participant's role in MAD~\cite{p03-li-2024, p11-Huang-2024, p16-Bai-2024, p28-chan-2024, s14-pham-2024, s17-du-2024, s18-chen-2025, s21-li-2024, s23-chen-2024, s28-xu-2023}. To ensure that diverse perspectives are represented in the debate, some approaches assign multiple \textit{personas} to each debater~\cite{p11-Huang-2024,p20-Jeptoo-2025, p28-chan-2024, s11-Yu-He-2024, s09-yang-2024}.

In addition to debaters, other roles have been defined and used in the literature to improve the structure of the debate. A common role is that of  \textit{summarizer}~\cite{p17-wang-2025, p20-Jeptoo-2025, p21-wang-2024,p26-li-2024, p28-chan-2024, s05-wang-2024,s11-Yu-He-2024}, an AI agent responsible for synthesizing the opinions expressed by debaters during the debate. In some cases, a \textit{judge}~\cite{p02-fang-2025,p04-liang-2024,p09-lim-2024, p20-Jeptoo-2025, p21-wang-2024, p26-li-2024, s05-wang-2024,s11-Yu-He-2024, s12-xiong-2023, s24-wang-2024} is included, an AI agent responsible for making a final decision based on the debate among the \textit{debaters}. We also found a few approaches incorporating a \textit{leader} (also called  \textit{moderator})~\cite{p11-Huang-2024, p17-wang-2025,s11-Yu-He-2024, s29-wang-2024} to oversee and guide the debate; a \textit{verifier} (also called  \textit{factuality checker})~\cite{s09-yang-2024, s27-Zhu-2023} who is in charge to fact-check the statements from the debaters; and an \textit{editor}~\cite{p26-li-2024} who is responsible to edit and generate the final response.

\textbf{Personas}: 
A \textit{participant} may have defined a specific persona, defining its \textit{background}, \textit{stance}, and/or \textit{personality traits}. 

\textit{Background} is usually domain-specific and represent the expertise and interests of the AI-agent (e.g. medical expert, data scientist, mathematician).   
These backgrounds are usually \textit{manually identified}, although we found an approach where backgrounds were \textit{automatically identified} based on the tasks to perform (i.e given a task that requires debate, it automatically generates a list of participants with different personas)~\cite{s29-wang-2024}.

A debater may also be given a predefined \textit{stance}. In debates involving two debaters, a common configuration involves one debater taking a positive stance and the other adopting a negative stance (usually named the ``angel" and the ``devil", respectively)~\cite{p04-liang-2024,p26-li-2024}.
In other formats, one debater is given the stance to challenge another debater. This AI agent is usually called the \textit{critic}~\cite{p02-fang-2025, p21-wang-2024}, a specialized debater whose task is to respond with counterarguments. 

Finally, one approach defined \textit{personality traits} of the AI agents, such as being more stubborn or suggestible~\cite{s18-chen-2025} to influence the outcome of the final agreement. 

\subsubsection{Interaction}
Under this category, we identified three sub-categories discussing the \textit{topology}, \textit{protocol} and \textit{format} of the debate.

\textbf{Topology}: 
Typically, the topology of the debates follow a \textit{fully connected} format, where every debater interacts with each other. Among\textit{ fully connected} topologies, a common setting involves two single AI agents participating in a \textit{bilateral} debate, such as those involving a \textit{debater} vs. a \textit{critic} or an \textit{angel} vs. \textit{devil}~\cite{p02-fang-2025,p04-liang-2024,p09-lim-2024, p26-li-2024}. 
Beyond these basic configurations, most approaches support an arbitrary number of debaters~\cite{p11-Huang-2024,p16-Bai-2024,p17-wang-2025,p20-Jeptoo-2025,p21-wang-2024,p28-chan-2024,s05-wang-2024,s09-yang-2024,s11-Yu-He-2024,s12-xiong-2023,s14-pham-2024,s17-du-2024, s23-chen-2024,s28-xu-2023}

However, this configuration may present scalability issues. To address this issue, we found one approach where debaters were \textit{divided into (smaller) groups} to fully debate internally among them, and engaging in reduced communication with members of the other groups~\cite{s24-wang-2024}.
With a similar simplification aim, other approaches introduced \textit{structured networks}, explicitly defining which debaters interact with whom, reducing the total number of interactions~\cite{p03-li-2024,s18-chen-2025, s29-wang-2024}. 


\textbf{Protocol}: The interaction among debaters can be \textit{sequential}, where debaters take turns expressing their opinions~\cite{p02-fang-2025,p04-liang-2024,p09-lim-2024,p11-Huang-2024, p28-chan-2024,s09-yang-2024,s11-Yu-He-2024, s27-Zhu-2023}, 
this is usually the case in the bilateral structures.

Other approaches use \textit{simultaneous} discussions, where multiple debaters express their opinions in rounds. Each debater shares their perspective in the first round, and in the following rounds they receive the opinions of the previous round from the other debaters, mitigating the risk of any bias caused by the order in which they debate~\cite{p03-li-2024,p11-Huang-2024,p16-Bai-2024,p20-Jeptoo-2025,p21-wang-2024,p26-li-2024,p28-chan-2024,s12-xiong-2023,s14-pham-2024,s17-du-2024,s18-chen-2025,s21-li-2024,s23-chen-2024,s24-wang-2024,s28-xu-2023,s29-wang-2024}. Interestingly, we found one \textit{hybrid} approach combining both strategies: starting with simultaneous talk in the first round and following with a sequential protocol for the remainder of the debate~\cite{s05-wang-2024}.

\textbf{Format}: 
The format in which debaters exchange information is usually \textit{natural language}. However, we found one approach where debaters exchanged their \textit{embedding} vectors to avoid potential risk of information loss~\cite{s14-pham-2024}

From those approaches who use natural language, we found that usually, the outputs from debaters are directly used as inputs for others during the debate (\textit{verbatim exchanges}). However, to reduce the number of tokens used, some approaches use \textit{summarized content} instead~\cite{p20-Jeptoo-2025, p28-chan-2024}. Some approaches include \textit{additional information} in the communication exchange, such as asking debaters to provide a \textit{confidence level} with their responses~\cite{p16-Bai-2024, s23-chen-2024}, or involving a special AI agent acting as a fact-checker and providing a \textit{factuality score}~\cite{s09-yang-2024}. 

\subsubsection{Type of agreement}

In some approaches, agreement is reached \textit{collectively} by the debaters themselves. Although the goal of most MAD approaches is to reach consensus, this might not always be the case even after a long debate. In these cases, the final collective agreement is obtained by \textit{majority vote}~\cite{p03-li-2024, p16-Bai-2024, p20-Jeptoo-2025, p28-chan-2024, s09-yang-2024, s14-pham-2024, s17-du-2024, s28-xu-2023}, \textit{weighted or scoring-based vote} based on the weighted votes obtained from the debaters~\cite{p11-Huang-2024, s23-chen-2024}, or \textit{averaging} the responses if they are numerical~\cite{p28-chan-2024, s09-yang-2024}.

Other family of approaches rely on a judge to determine the most convincing argument~\cite{p02-fang-2025, p04-liang-2024, p09-lim-2024, s05-wang-2024, s11-Yu-He-2024, s12-xiong-2023}. This \textit{judge-based agreement} is especially common again in bilateral structures. The judge is usually given access to the entire debate, but we found some approaches where only a summary provided by a summarizer was given~\cite{p09-lim-2024}. In some cases, the judge’s role is limited to deciding when consensus has been reached, while the summarizer or editor produces the final answer~\cite{p17-wang-2025, p21-wang-2024}. 
In one instance, we found an hybrid approach, where the judge is only used to decide if a tie has been reached after the votes of the debaters~\cite{s24-wang-2024}.

Together, these dimensions illustrate the richness and diversity of methodologies in MAD approaches, highlighting both the flexibility of the paradigm and the patterns that shape its current landscape.

\section{MAD strategies applied to RE}

\subsection{Design}

To answer RQ2, we conducted an empirical evaluation of representative MAD within the scope of a well-established RE task: binary classification of requirements. Given a set of natural language requirements $R = \{r_1, r_2, \dots, r_n\}$, the objective is to assign each $r_i$ to one of two mutually exclusive categories: functional (F) or non-functional (NF). 

We used as a baseline a single-agent classification pipeline as proposed by Ronanki et al.~\cite{Ronanki2024}. In line with their study, we used prompt-based interaction with an LLM agent, preserving the \textit{Persona} prompt pattern proposed in their original study:

\begin{tcolorbox}[
colback=gray!5!white,
colframe=gray!75!black,
colbacktitle=blue!10!white,
coltitle=black,
title=\textbf{Baseline} (\textit{System prompt}),
boxrule=0.8pt,
left=0.5mm, right=0.5mm, top=0.5mm, bottom=0.5mm
]
Act as a requirements engineering domain expert and classify the given list of requirements into functional (F) and non-functional requirements (NF).
\end{tcolorbox}

\begin{figure}[t]
  \centering
  \includegraphics[width=\columnwidth]{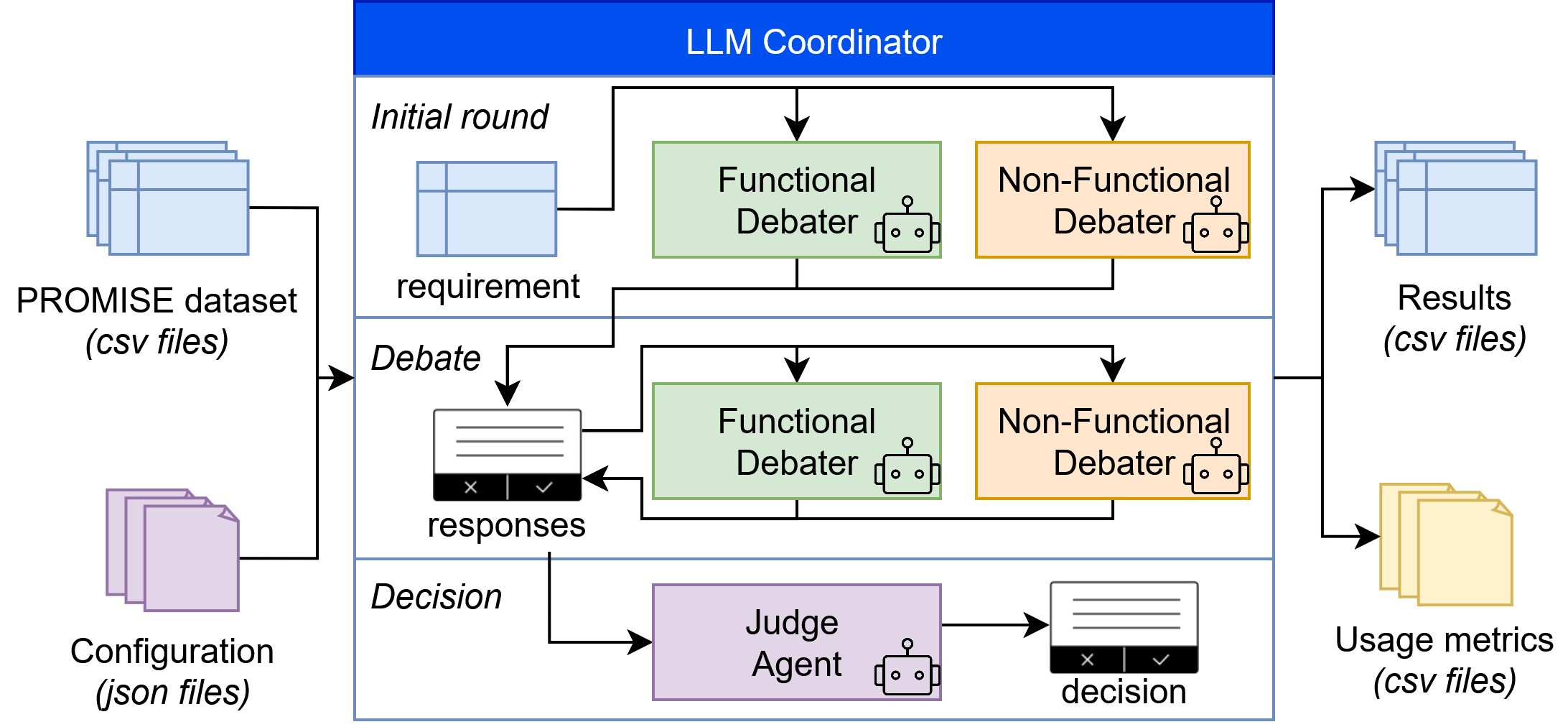}
  \caption{MAD architecture and interaction workflow.}
  \label{fig:architecture}
\end{figure}

To assess the potential of MAD strategies, we designed a debate-based classification system, illustrated in Figure~\ref{fig:architecture}. The MAD setting conducted within the context of this research preview consists of three participants, consisting of two debaters and one judge. To adapt these roles to our task, we refined their stance to identify the following participants: (\textit{i}) a \textit{Functional debater}, arguing why a given requirement $r_i$ is functional; (\textit{ii}) a \textit{Non-Functional debater}, arguing why a given requirement $r_i$ is non-functional; and (\textit{iii}) a \textit{Judge}, responsible for making the final classification decision $f : R \rightarrow \{F, NF\}$.

\begin{tcolorbox}[
colback=gray!5!white,
colframe=gray!75!black,
colbacktitle=green!10!white,
coltitle=black,
title=\textbf{MAD - Functional debater} (\textit{System prompt}),
boxrule=0.8pt,
left=0.5mm, right=0.5mm, top=0.5mm, bottom=0.5mm
]
You are a debater arguing that the received requirement is functional (F).
\end{tcolorbox}

\begin{tcolorbox}[
colback=gray!5!white,
colframe=gray!75!black,
colbacktitle=orange!10!white,
coltitle=black,
title=\textbf{MAD - Non-Functional debater}  (\textit{System prompt}),
boxrule=0.8pt,
left=0.5mm, right=0.5mm, top=0.5mm, bottom=0.5mm
]
You are a debater arguing that the received requirement is non-functional (NF).
\end{tcolorbox}

\begin{tcolorbox}[
colback=gray!5!white,
colframe=gray!75!black,
colbacktitle=blue!10!white,
coltitle=black,
title=\textbf{MAD - Judge}  (\textit{System prompt}),
boxrule=0.8pt,
left=0.5mm, right=0.5mm, top=0.5mm, bottom=0.5mm
]
You are a moderator. There will be two debaters involved in a debate competition. They will present their answers and discuss their perspectives on whether a given requirement should be classified as functional (F) or non-functional (NF). At the end of each round, you will evaluate their answers and decide which classification is more appropriate.
\end{tcolorbox}

Interaction within the MAD strategy is organized using a structured network of simultaneous participation of each debater, which we defined as follows:

\begin{enumerate}
    \item A given requirement $r_i$ (either \textit{Functional} or \textit{Non-Functional}) is sent to both debaters for classification.
    \item In the \textit{basic} setting ($n = 0$), both debaters independently generate their arguments based solely on the requirement. These verbatim responses are then submitted to the \textit{Judge}, which produces the final classification $\{F,NF\}$.
    \item In the \textit{iterative} variant ($n > 0$), a discussion mechanism between debaters is introduced:
    \begin{itemize}
        \item For each iteration $t \in \{1, \dots,n\}$, each debater receives their opponent's latest verbatim argument, generating a rebuttal or refinement of their position.
        \item After $n$ iterations, the \textit{Judge} evaluates the entire discussion transcript to generate a judge-based agreement, determining the final classification.
    \end{itemize}
\end{enumerate}

The parameter $n$ thus controls the depth of the debate and can be tuned to explore the effect of iterative refinement and feedback exchange between debaters.

All agents in our experiments were implemented using the OpenAI Assistants API\footnote{https://platform.openai.com/docs/assistants/}, with GPT-4o as the underlying model. We fixed the temperature to $\tau = 0.0$ across all experiments to reduce non-determinism and improve the reproducibility of responses, reducing stochastic variability in text generation from debaters and prediction class from the judge.

For our empirical evaluation, we used the PROMISE dataset~\cite{ClelandHuang2007} (replicating the evaluation by Ronanki et al.~\cite{Ronanki2024}), which consists of 621 labeled requirements: 253 functional and 368 non-functional. We analysed three settings: (\textit{i}) the single-agent baseline; (\textit{ii}) the MAD strategy with $n=0$ (i.e., no interaction between debaters); and (\textit{iii}) the MAD strategy with $n=1$ (i.e., one round of interaction between debaters).

Our evaluation focuses on two quality dimensions. First, \textit{correctness}, which we measure using standard classification metrics: accuracy, precision, recall, and F1-score. Second, \textit{cost-efficiency}, which we assess based on total token usage, including input (\textit{I-Tokens}) and output (\textit{O-tokens}), execution time, and monetary cost. The code, datasets and experimental results are available in GitHub\footnote{https://github.com/nlp4se/llm-discussion/}.

\subsection{Evaluation}

Table~\ref{tab:results} reports the results for the correctness analysis of the three sets of experiments, including class-specific and average metrics (weighted by class support). MAD strategies consistently outperform the single-agent baseline across all metrics. Notably, MAD without interactions ($n=0$) already yields a substantial improvement in F1-score over the baseline, from $0.726$ to $0.835$ ($+0.109$). Introducing a single round of interaction ($n=1$) barely improves the overall results, reaching an accuracy of $0.825$ ($+0.009$) and an F1-score of $0.841$ ($+0.006$). Within the context of this research preview, given the reduced improvement on correctness metrics and the significant increase of resource consumption (see discussion on Table~\ref{tab:cost-efficiency}), we decided not to conduct experiments with $n>1$.

\begin{table}[h]
\caption{Correctness analysis}
\label{tab:results}
\begin{tabularx}{\linewidth}{@{}lXcccc@{}}
\toprule
\textbf{Method} & \textbf{Class} & \textbf{Accuracy} & \textbf{Precision} & \textbf{Recall} & \textbf{F1-score} \\ \midrule
\multirow{3}{*}{\textbf{Baseline}}  & \textbf{F}     & --     & 0.613 & 0.893 & 0.727 \\
                                    & \textbf{NF}    & --     & 0.893 & 0.611 & 0.726 \\
                                    & \textbf{Total} & 0.726  & 0.779 & 0.726 & 0.726 \\ \midrule
\multirow{3}{*}{\textbf{MAD (n=0)}} & \textbf{F}     & --     & 0.701 & 0.957 & 0.809 \\
                                    & \textbf{NF}    & --     & 0.960 & 0.720 & 0.823 \\
                                    & \textbf{Total} & 0.816  & 0.855 & 0.816 & 0.835 \\ \midrule
\multirow{3}{*}{\textbf{MAD (n=1)}} & \textbf{F}     & --     & 0.713 & 0.953 & 0.816 \\
                                    & \textbf{NF}    & --     & 0.958 & 0.737 & 0.833 \\
                                    & \textbf{Total} & 0.825  & 0.858 & 0.824 & 0.841 \\ \bottomrule
\end{tabularx}
\end{table}

Focusing on class-specific improvements, the most substantial gains are observed for NF requirements. The F1-score for NF increases from $0.726$ in the baseline to $0.833$ with MAD at $n = 1$, a notable gain of $+0.107$. This improvement is primarily driven by a significant boost in recall, which rises from $0.611$ to $0.737$ ($+0.126$). While MAD strategies help reduce the imbalance between precision and recall across classes, the overall pattern remains consistent across configurations: functional requirements exhibit high recall ($>0.950$), and non-functional requirements maintain very high precision ($>0.950$). Conversely, the precision for functional requirements and the recall for non-functional requirements, while still comparatively lower ($>0.700$), show substantial improvement over their respective baseline values ($<0.620$).

Table~\ref{tab:cost-efficiency} reports the results for the cost-efficiency analysis. As expected, the adoption of MAD strategies results in a substantial increase in resource consumption. Compared to our baseline, MAD with $n=0$ requires approximately $9.5\times$ more tokens and $3.4\times$ more time, with an estimated cost increase of over \textbf{16$\times$} (from 0.43\,€ to 6.98\,€). This trend becomes even more pronounced for $n=1$, which doubles the token usage and cost again — representing a $33.5\times$ increase in cost compared to the baseline and a $6.7\times$ increase in time. Overall, while MAD strategies yield measurable improvements in classification performance, particularly at $n=0$, the marginal performance gain observed at $n=1$ (e.g., $+0.006$ F1-score) comes at the expense of doubling both the cost and runtime.

\begin{table}[h]
\caption{Cost-efficiency analysis}
\label{tab:cost-efficiency}
\begin{tabular}
{@{}lrrrrr@{}}
\toprule
\textbf{Method}                                     & \textbf{I-Tokens} & \textbf{O-Tokens} & \textbf{Tokens} & \textbf{Cost} & \textbf{Time} \\ \midrule
\textbf{Baseline} & 152,145 & 8,694 & 160,839 & 0.43 € & 1.9 h \\ 
\textbf{MAD (n=0)} & 1,068,517 & 453,126 & 1,521,643 & 6.98 € & 6.4 h\\ 
\textbf{MAD (n=1)} & 2,137,034 & 906,252 & 3,043,286 & 14.41 € & 12.8 h \\ 
\bottomrule
\end{tabular}
\end{table}

\subsection*{Statistical Significance Analysis}
To assess whether the observed performance improvements between classifiers are statistically significant, we applied McNemar’s test on the prediction outcomes. This non-parametric test is well-suited for comparing two classifiers evaluated on the same dataset, particularly in binary classification settings, as it evaluates whether the disagreement in their predictions is significant beyond chance.

We computed the test using the predictions over the 621 requirements in the test set. The contingency tables and resulting $p$-values are presented in Table~\ref{tab:mcnemar}. In both comparisons ---MAD ($n=0$) vs. baseline and MAD ($n=1$) vs. baseline--- the differences were found to be statistically significant, with $p < 0.001$. This confirms that the observed gains in classification performance are statistically significant and unlikely to be due to random variation.

\begin{table}[h]
\caption{McNemar’s test: Comparison of MAD strategies vs baseline}
\label{tab:mcnemar}
\centering
\begin{tabularx}
{\linewidth}{@{}Xcccc@{}}
\toprule
\textbf{Comparison} & \textbf{b} & \textbf{c} &  &\textbf{p-value} \\
\midrule
\textbf{MAD (n=0) vs Baseline} & 71 & 15 &  &\multicolumn{1}{l}{$3.0 \times 10^{-9}$} \\
\textbf{MAD (n=1) vs Baseline} & 75 & 14 & &\multicolumn{1}{l}{$2.0 \times 10^{-10}$} \\
\bottomrule
\end{tabularx}
\vspace{1mm}

\footnotesize
\centerline{\textit{b}: cases where MAD is correct and baseline is wrong.}
\centerline{\textit{c}: cases where baseline is correct and MAD is wrong.}
\end{table}

\subsection{Discussion}
The publication years of the papers indicate MAD is a new and growing field. Despite its potential, we did not find any contribution applying MAD in RE nor in the Software Engineering tasks. The analysis of 25 publications revealed a diverse landscape of approaches, each with unique characteristics regarding participant \textit{roles} and \textit{personas}; the \textit{topology}, \textit{protocol} and \textit{format} of the debate; and the \textit{agreement} mechanisms applied. While some approaches prioritize nuanced debates through detailed \textit{personas}, multiple \textit{roles} and \textit{fully connected} interactions, others focus on efficiency and scalability through \textit{summarized content} and \textit{structured networks}. 

Our experiments illustrate how MAD strategies can effectively be applied to improve the correctness of traditional RE tasks such as binary requirements classification. However, evaluating the trade-off between accuracy gains and computational overhead is critical not only for practical deployment but also from a sustainability standpoint, given the energy and cost implications of scaling such systems. This raises concerns about the scalability and sustainability of multi-agent reasoning approaches, especially for large-scale deployments or environments with limited computational resources. As shown in Table~\ref{tab:cost-efficiency}, increasing the number of debate iterations significantly increases both token consumption and inference time. However, the corresponding improvements in quality metrics (Table~\ref{tab:results}) are marginal. Given this imbalance between performance gains and resource demands, we limited our research preview to a single round of discussion ($n=1$), leaving deeper multi-round reasoning for future work. Additionally, assessing variability for different MAD strategies, prompts, temperature values and inference runs remains also to be explored, opening the perspective to higher improvements in correctness metrics and optimized token consumption.

\section{Research plan}

\begin{figure*}[htbp]
  \centering
  \includegraphics[width=0.7\textwidth]{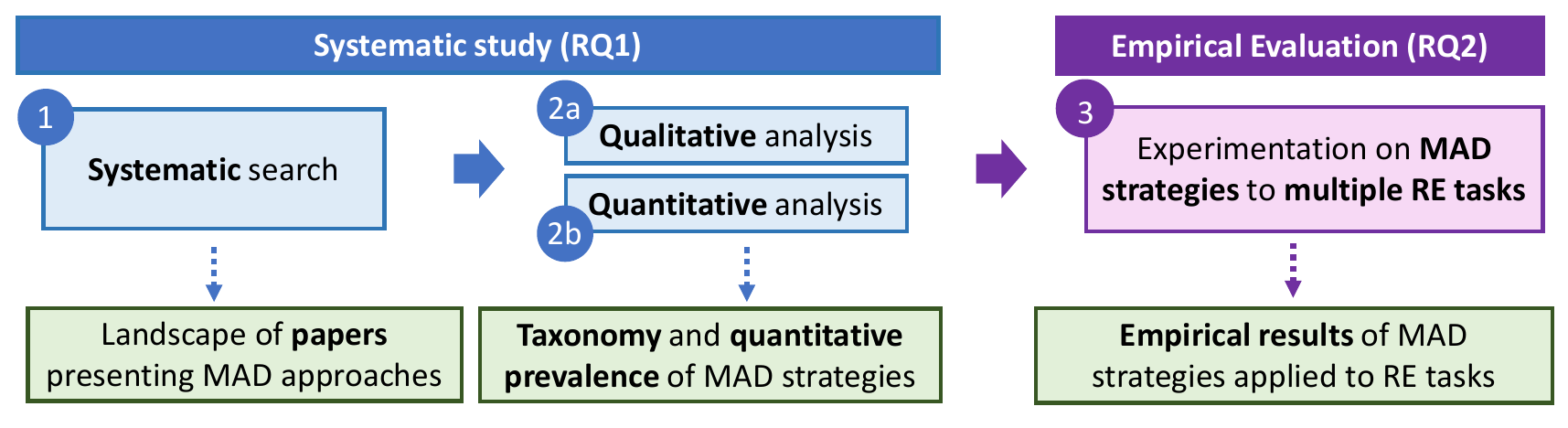}
  \caption{Research plan}
  \label{fig:research_plan}
\end{figure*}

The presented research focuses on the analysis of the MAD strategies proposed in the literature and their key characteristics (RQ1) and to which degree these MAD strategies can be applied to RE tasks (RQ2). A general overview of the research plan we envision to fully answer these research question is depicted in Figure~\ref{fig:research_plan}.

\label{sec:search-analysis}

\textit{1. Systematic search: } We plan to finalize the systematic mapping study, following the search methodology presented in Section~\ref{sec:search-methodology}, by iteratively conducting both backward and forward snowballing until saturation is reached. The outcome of this activity will be the landscape of papers in the literature presenting MAD approaches across multiple domains.

\textit{2a. Qualitative analysis: } We will conclude the qualitative analysis by applying the content analysis outlined in Section ~\ref{sec:search-analysis} to the newly identified papers from the completed systematic search conducted in activity (1). The goal of this activity is to refine the existing taxonomy by ensuring that all relevant strategies are accurately represented and properly categorized.

\textit{2b. Quantitative analysis: } We will perform a comprehensive quantitative analysis of the MAD strategies identified in the literature. This analysis will include evaluating the frequency of the occurrence of each strategy, its distribution in various research domains, and an examination of the results obtained from the different types of tasks on which they have been tested.

The outcomes of activities (2a) and (2b) will be a refined taxonomy with a quantitative analysis of the different strategies.

\textit{3. Experimentation with MAD strategies for multiple RE tasks}. We plan to conduct a comprehensive experimentation on the most promising MAD strategies as identified in activities (2a) and (2b), by applying them to a variety of RE tasks, including what are known as hairy RE tasks~\cite{Berry2021}, for which maximizing recall is particularly critical. Some of the RE tasks we envision to address are requirements classification, requirements traceability and ambiguity detection, among others. To conduct these experiments, we will use well-known publicly available datasets, including datasets from the RE Open Data Initiative\footnote{https://zenodo.org/communities/re-data/}.
Besides the accuracy and cost, we plan to measure other critical aspects like the trustworthiness of these MAD strategies through metrics related to the transparency, explainability and robustness from Trustworthy AI research~\cite{chander2025}. Additionally, we will conduct empirical evaluations to identify potential biases in LLMs when performing RE tasks and assess how MAD strategies may help mitigate them.  

The outcome of this activity will be a comprehensive evaluation of the most effective MAD strategies across different RE tasks. 

\section{Potential risks and limitations}
In this section, we discuss the potential risks and limitations we identified in our research plan and methodology, along with the mitigation actions we envisioned to minimize their impact.

\textit{Rapid evolution of the field}: MAD is a novel field that emerged in 2023 and, with the fast pace advancements of AI models, it is likely to also evolve rapidly. This dynamic nature may limit the applicability of the results of both quantitative and qualitative analyses in the future. To address this threat, we employed Wohlin’s systematic methodology for conducting systematic literature studies~\cite{Wohlin2014}, which facilitates iterative refinement of the literature through backward and forward snowballing.  Furthermore the full experimentation pipeline will be designed for easy re-execution to allow replication with multiple technologies and adapt to model evolution.

\textit{Data extraction for the development of the taxonomy}: The taxonomy development depends on manual content analysis. To address potential threats to the data extraction process, we employed a structured coding strategy inspired by thematic analysis~\cite{Cruzes2011}.  

\textit{Generalizability of the results}: In the current evaluation, we have chosen a particular MAD strategy, RE task and dataset, using a single LLM, which clearly limits the  generalizability of the results. This evaluation has been conducted for illustrative purposes and we plan to incorporate multiple MAD strategies, RE tasks and datasets in subsequent evaluations. Furthermore, we plan to use other LLMs, including reasoning models such as GPT-o1.

\textit{Lack of high-quality datasets}: The envisioned empirical evaluation require the existence of high-quality datasets that can be used as ground-truth for automating the experiments. To address this issue, we plan to use well-known publicly available datasets that have already been successfully used in the literature. However, certain requirements engineering (RE) tasks present a significant challenge in obtaining such ground truth (e.g. requirements elicitation). In this regard, our research results will be limited to RE tasks where a ground truth is available. 

\section{Conclusions}
In this research preview we presented the initial results of our ongoing study that explores the potential of using MAD strategies to enhance the accuracy of AI agents in RE tasks. 

By means of a systematic mapping, we analysed the existing MAD strategies proposed in the literature and developed a taxonomy that categorizes their characteristics, considering three primary categories: the \textit{participants} involved, the \textit{interactions} during the debate, and the strategies for reaching an \textit{agreement}. We argue that the presented taxonomy not only offers a lens to understand current practices but also serves as a foundation for guiding future research.

By leveraging this taxonomy, we selected a specific MAD strategy to conduct an empirical evaluation over the task of binary requirements classification (functional vs non-functional). Our findings suggest that MAD is a promising approach for improving AI agents' performance in RE tasks, albeit at a higher cost, by utilizing collaborative and iterative debate mechanisms among AI agents.

From our systematic study, we also concluded that no MAD strategy has previously been applied to RE tasks. This research aims at providing a comprehensive understanding of MAD strategies and their applicability to RE tasks. As future work, we plan to execute the outlined research plan and validate the effectiveness of different MAD strategies in enhancing AI agents' performance across multiple and more complex RE tasks.

\section*{Acknowledgments}
This work has been supported by funding from the HIVEMIND project – Horizon Europe call HORIZON-CL4-2024-DIGITAL-EMERGING-01 under Grant Agreement Number 101189745; and  by the Spanish Ministerio de Ciencia e Innovación under project funding scheme PID2020-117191RB-I00 / AEI/10.13039/501100011033.




\bibliographystyle{IEEEtran}
\bibliography{references}
\end{document}